\documentclass{desyproc}

\usepackage{cite}

\graphicspath{{figs/}}

\input{paperdef}

{
\newcommand{\sixoo}{60 \ifb}
\newcommand{\sixooeff}{60 \ifb\,eff$\times2$}
\newcommand{\sixooo}{600 \ifb}
\newcommand{\sixoooeff}{600 \ifb\,eff$\times2$}
\newcommand{\Mhmax}{M_h^{\rm max}}

\begin{document}

\thispagestyle{empty}
\setcounter{page}{0}
\def\thefootnote{\fnsymbol{footnote}}

\begin{flushright}
IPPP/09/71\\
DCPT/09/142\\
arXiv:0909.4665 [hep-ph]
\end{flushright}

\vspace{1cm}

\begin{center}

{\fontsize{15}{1} 
\sc {\bf BSM Higgs Studies at the LHC\\[.5em] 
in the Forward Proton Mode}}
\footnote{talk given by S.H.\ and V.A.K.\ at the {\em EDS\,09}, 
July 2009, CERN}

\vspace{1cm}

{\sc 
S.~Heinemeyer$^1$%
\footnote{
email: Sven.Heinemeyer@cern.ch
}%
, V.A.~Khoze$^2$%
\footnote{
email: v.a.khoze@durham.ac.uk
}%
, M.G.~Ryskin$^3$%
\footnote{
email: ryskin@thd.pnpi.spb.ru
}%
, M.~Ta\v{s}evsk\'{y}$^4$%
\footnote{
email: tasevsky@mail.cern.ch
}%
~and G.~Weiglein$^2$%
\footnote{
email: Georg.Weiglein@durham.ac.uk
}
}

\vspace*{1cm}

$^1$ Instituto de F\'isica de Cantabria (CSIC-UC), Santander,  Spain

\vspace*{0.1cm}

$^2$ IPPP, University of Durham, Durham DH1~3LE, UK

\vspace*{0.1cm}

$^3$ Petersburg Nuclear Physics Institute, Gatchina, St. Petersburg, 188300, 
Russia

\vspace*{0.1cm}

$^4$ Institute of Physics of the ASCR, 
Na Slovance 2, 18221 Prague 8, Czech Republic
 
\end{center}

\vspace*{0.2cm}

\begin{center} {\bf Abstract} \end{center}
The prospects for central exclusive diffractive (CED) production of 
BSM Higgs bosons at the LHC are reviewed.
This comprises the production of MSSM and 4th generation Higgs bosons.
The sensitivity of the searches in the forward proton
mode for the Higgs bosons as well as the possibility of a coupling
structure determination are briefly discussed.

\def\thefootnote{\arabic{footnote}}
\setcounter{footnote}{0}

\newpage

\title{BSM Higgs Studies at the LHC in the Forward Proton Mode}

\author{{\slshape S. Heinemeyer$^1$, V.A. Khoze$^{2}$,
M.G. Ryskin$^{3}$, M. Ta\v{s}evsk\'{y}$^4$ and G. Weiglein$^2$ }\\[1ex]
$^1$Instituto de F\'isica de Cantabria (CSIC), Santander, Spain,\\ 
$^2$IPPP, Department of Physics, Durham University, Durham DH1 3LE, U.K.,\\ 
$^3$Petersburg Nuclear Physics Institute, Gatchina, St. Petersburg, 188300, 
Russia,\\ 
$^4$Institute of Physics of the ASCR, 
Na Slovance 2, 18221 Prague 8, Czech Republic}

\contribID{heinemeyer\_sven}

\desyproc{DESY-PROC-2009-xx}
\acronym{EDS'09} 
\doi  

\maketitle

\begin{abstract}
The prospects for central exclusive diffractive (CED) production of 
BSM Higgs bosons at the LHC are reviewed.
This comprises the production of MSSM and 4th generation Higgs bosons.
The sensitivity of the searches in the forward proton
mode for the Higgs bosons as well as the possibility of a coupling
structure determination are briefly discussed.

\end{abstract}


\section{Introduction}

In recent years, there has been a growing interest
in the possibility to complement the standard LHC physics menu by
installing  near-beam proton detectors in the LHC tunnel. 
Projects to install the proton detectors at 220~m and 420~m from the
interaction points are now under review inside ATLAS and
CMS~\cite{DKMOR,CMS-Totem,FP420,AFP}. 
The combined detection of both outgoing protons and the centrally
produced system gives access to a rich program
of studies of QCD, electroweak and BSM physics, see for
instance \cite{FP420,INC}. Importantly,
these measurements will
provide valuable information on the Higgs sector of MSSM and other popular
BSM scenarios, see \cite{KKMRext,HKRSTW,CLP,fghpp,ismd,tripl}.

As it is well known,
many models of new physics require an extended
Higgs sector. The most popular extension of the SM 
is the 
MSSM, where the Higgs sector consists of
five physical states. At lowest order the MSSM Higgs
sector is $\cp$-conserving, containing two $\cp$-even
bosons, the lighter $h$ and the heavier $H$, 
a  $\cp$-odd boson, $A$, and the charged
bosons $H^\pm$. It can be specified
 in terms of the gauge couplings, the ratio of the two vacuum
expectation values, $\tb \equiv v_2/v_1$, and the mass of the $A$
boson, $\MA$.
The Higgs phenomenology 
in the  MSSM is strongly affected by  higher-order
corrections (see \cite{reviews} for reviews).

Another  very simple example of physics beyond the SM is a
model which extends the SM by a fourth generation of heavy fermions
(SM4), see, for instance,~\cite{extra-gen-review}. 
Here the masses of the 4th generation quarks and leptons
are assumed to be (much) heavier than the mass of the top-quark.
In this case, the effective coupling of the Higgs boson
to two gluons is three times larger than in the SM, and all branching
ratios change correspondingly.

Proving that a detected new state is, indeed, a Higgs boson and
distinguishing the Higgs boson(s) of the SM, the SM4 or the MSSM from
the states of other theories will be far from trivial.
In particular, it will be of utmost importance 
to determine the spin and
$\cp$ properties of a new state and to measure precisely its mass, width
and couplings.

The CED processes are of the form~
~$pp\to p \oplus H \oplus p$, where the $\oplus$ signs denote
 large rapidity gaps on either side of the   centrally produced state.
 If the outgoing protons remain intact and scatter
through small
angles then, to a very good approximation, the primary di-gluon
system obeys a $J_z=0$, $\cp$-even selection rule~\cite{KMRmm}. 
Here $J_z$ is the projection of the total
angular momentum along the proton beam. This 
permits a clean determination of the quantum numbers of the
observed resonance which  will be dominantly produced in a $0^+$ 
state. Furthermore, because the process is exclusive, the
proton  energy losses are directly related to the central mass,
allowing a potentially excellent mass resolution,
irrespective of the decay channel. The CED
processes allow in principle all the main
Higgs decay modes, $b \bar b$,  $WW$  and $\tau\tau$, to be
observed in the same production channel. In particular, a unique
possibility opens up to study the Higgs Yukawa coupling to bottom
quarks, which, as it is well known, may be difficult
to access in other search channels at the LHC. Here it should be kept in
mind that access to the bottom Yukawa coupling will be crucial as an
input also for the determination of 
Higgs couplings to other particles~\cite{HcoupLHCSM,HcoupLHC120}.

Within the MSSM, CED production is even more appealing than in the SM. 
The lightest MSSM Higgs boson coupling to $b \bar b$ and $\tau\tau$ 
can be strongly enhanced for large values of
$\tb$ and relatively small $\MA$. On the other hand, for
larger values of $\MA$ the branching ratio of $H \to b \bar b$ 
is much larger
than for a SM Higgs of the same mass. As a consequence, CED 
$H \to b \bar b$ production can be studied in the MSSM up
to much higher masses than in the SM case.

Here we briefly review the analysis of~\cite{HKRSTW} 
where a detailed study of the CED MSSM Higgs
production was performed (see also \citeres{KKMRext,CLP,otherMSSM} for
other CED studies in the MSSM). This is updated  by taking into account 
recent theoretical developments in background evaluation~\cite{shuv,screen}
and using an improved version~\cite{FH2.6.2} of the code
{\tt FeynHiggs}~\cite{feynhiggs} employed for the cross section and
decay width calculations. The regions excluded by LEP and Tevatron Higgs
searches are evaluated with {\tt HiggsBounds}~\cite{higgsbounds}. 
These improvements are applied for the CED production of MSSM Higgs
bosons~\cite{HKRSTW} in the $\Mhmax$
benchmark scenario (defined in \cite{benchmark}), and in the SM4.


\section{Signal and background rates and experimental aspects}
\label{sec:backgrounds}

The Higgs signal and background cross sections can be 
approximated  by the simple formulae given in~\cite{KKMRext,HKRSTW}.
For CED production of the MSSM $h, H$-bosons the cross section 
$\si^{\rm excl}$ is 
\begin{equation}
\si^{\rm excl} \, \mbox{BR}^{\rm MSSM} =3 \, {\rm fb} 
  \left(\frac{136}{16+M}\right)^{3.3}
  \left(\frac{120}M\right)^3 
  \frac{\Ga(h/H \to gg)}{0.25\mev} \,\mbox{BR}^{\rm MSSM},
\label{eq1}
\end{equation}
where the gluonic width $\Ga(h/H\to gg)$ and the 
branching ratios for the various MSSM channels,
$\mbox{BR}^{\rm MSSM}$, are calculated with \fh {\tt 2.6.2}~\cite{FH2.6.2}.
The mass $M$ (in GeV) denotes either $\Mh$ or $\MH$.
The normalization is fixed  at $M=120 \gev$, where
 $\si^{\rm excl} =3$~fb for $\Ga(\HSM \to gg)=0.25 \mev$. 
In \citere{KKMRext,HKRSTW} the uncertainty in the prediction for the CED
cross sections was estimated to be below a factor of $\sim 2.5$. 
According to \cite{DKMOR,HKRSTW,shuv,krs2},
the overall background to the $0^+$ Higgs signal in the
$b \bar b$ mode can be approximated by
\BE
{\rm d}\si^B / {\rm d} M
\approx 0.5 \, {\rm fb/GeV} \left[
A (120/M)^6 + 1/2 \, C (120/M)^8
                                  \right] ~,
\label{eq:backbb}
\EE
with $A=0.92$ and $C=C_{\mathrm {NLO}}=0.48 - 0.12 \times (\ln(M/120))$.
The expression (\ref{eq:backbb}) holds for a mass window $\Delta M=4-5\gev$
and summarizes several types of
backgrounds: the prolific
$gg^{PP}\to gg$ subprocess can mimic $b\bar b$ production due
to the misidentification of the gluons as $b$ jets;
an admixture of $|J_z|=2$ production;
the radiative $gg^{PP}\to b\bar b g$ background;
due to the non-zero $b$-quark
mass there is also a contribution to the $J_z=0$ 
cross section of order $\mb^2/E_T^2$.
The first term in the square brackets corresponds
to the first three background sources~\cite{HKRSTW}, evaluated 
for $P_{g/b}=1.3 \%$, where $P_{g/b}$ is 
the probability to misidentify a gluon as a $b$-jet for a $b$-tagging
efficiency of 60\%.
The second term describes the background associated with bottom-mass
terms in the Born amplitude.
The NLO correction suppresses this contribution
by a factor of about 2, or more for larger masses~\cite{shuv}.

The main experimental challenge of running at high luminosity,\
$10^{34} \, {\rm cm}^{-2} \, {\rm s}^{-1}$, 
is the effect of pile-up, which can generate fake signal
events within the
acceptances of the proton detectors
as a result of the coincidence of two or more separate
interactions in the same bunch crossing, 
see \cite{CMS-Totem,FP420,HKRSTW,CLP} for details. 
Fortunately, as established in \cite{CLP}, the pile-up can be brought
under control by using time-of-flight vertexing and  cuts on the number
of charged tracks. Also in the analysis of \cite{HKRSTW}
the event selections and cuts were 
imposed such as to maximally reduce the pile-up background. 
Based on the anticipated improvements for a reduction of the 
overlap backgrounds down to a tolerable level, in the numerical
studies in \cite{CMS-Totem,HKRSTW} and in the new results below the pile-up
effects were not included.

At nominal LHC optics, proton taggers positioned at a distance
$\pm 420$~m from the interaction points of ATLAS and CMS will allow
a coverage of the proton
fractional momentum loss $\xi$ in the range 0.002--0.02,
with an acceptance of around 30\% for a centrally produced system
with a mass around $120 \gev$.
A combination with the foreseen proton detectors at
$\pm 220$~m~\cite{AFP,totemRP220} would enlarge
the $\xi$ range up to 0.2.
This would be especially beneficial because of
the increasing acceptance for higher masses \cite{HKRSTW}.
The main selection criteria for $h,H \to b \bar b$ are either two
$b$-tagged jets or two jets with at least one $b$-hadron decaying
into a muon. 
Details on the corresponding selection cuts and triggers for $b\bar b$, 
$WW$ and $\tau\tau$ channels can be found
in \cite{HKRSTW,CMS-Totem,cox1}. Following  \cite{HKRSTW} we
consider four luminosity
scenarios: ``\sixoo'' and ``\sixooo'' refer to running at low and high 
instantaneous luminosity, respectively, using conservative assumptions
for the signal rates and the experimental
sensitivities; possible improvements of both theory and experiment 
could lead to the scenarios where the
event rates are higher by a factor of 2, denoted as ``\sixooeff'' and
``\sixoooeff''.


\section{Updated sensitivities  for CED production
of the $\cp$-even MSSM Higgs bosons}
\label{sec:discovery}


Below we extend the analysis
of the  CED production  of $H \to b \bar b$ and  $H \to \tau \tau $
carried out in \cite{HKRSTW} and consider
the $\Mhmax$ benchmark scenario of \cite{benchmark}.
The improvements consist of the incorporation of the one-loop
corrections
to the mass-suppressed background~\cite{shuv} and in employing an
updated version of \fh~\cite{FH2.6.2,feynhiggs} for the cross section and
decay width calculations. Furthermore we now also display the limits in
the $\MA$--$\tb$ planes obtained from Higgs-boson searches at the
Tevatron. For the latter we employed the new code {\tt HiggsBounds}, see 
\cite{higgsbounds} (where also the list of CDF and D0 references for the
incorporated exclusion limits can be found). 
 
\begin{figure}[htb!]
\vspace{-1em}
\includegraphics[width=.495\textwidth,height=4.5cm]
{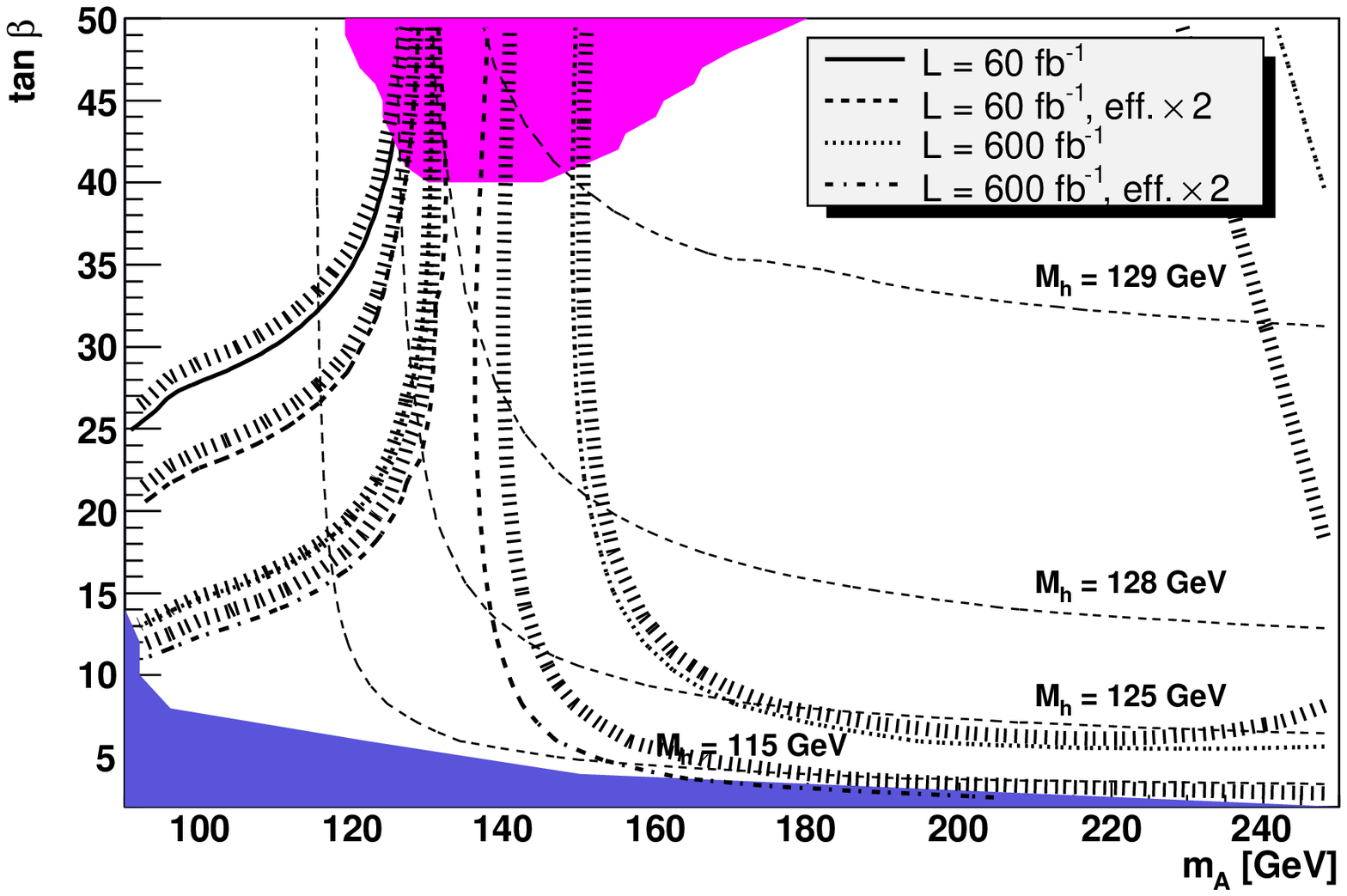}
\includegraphics[width=.495\textwidth,height=4.5cm]
{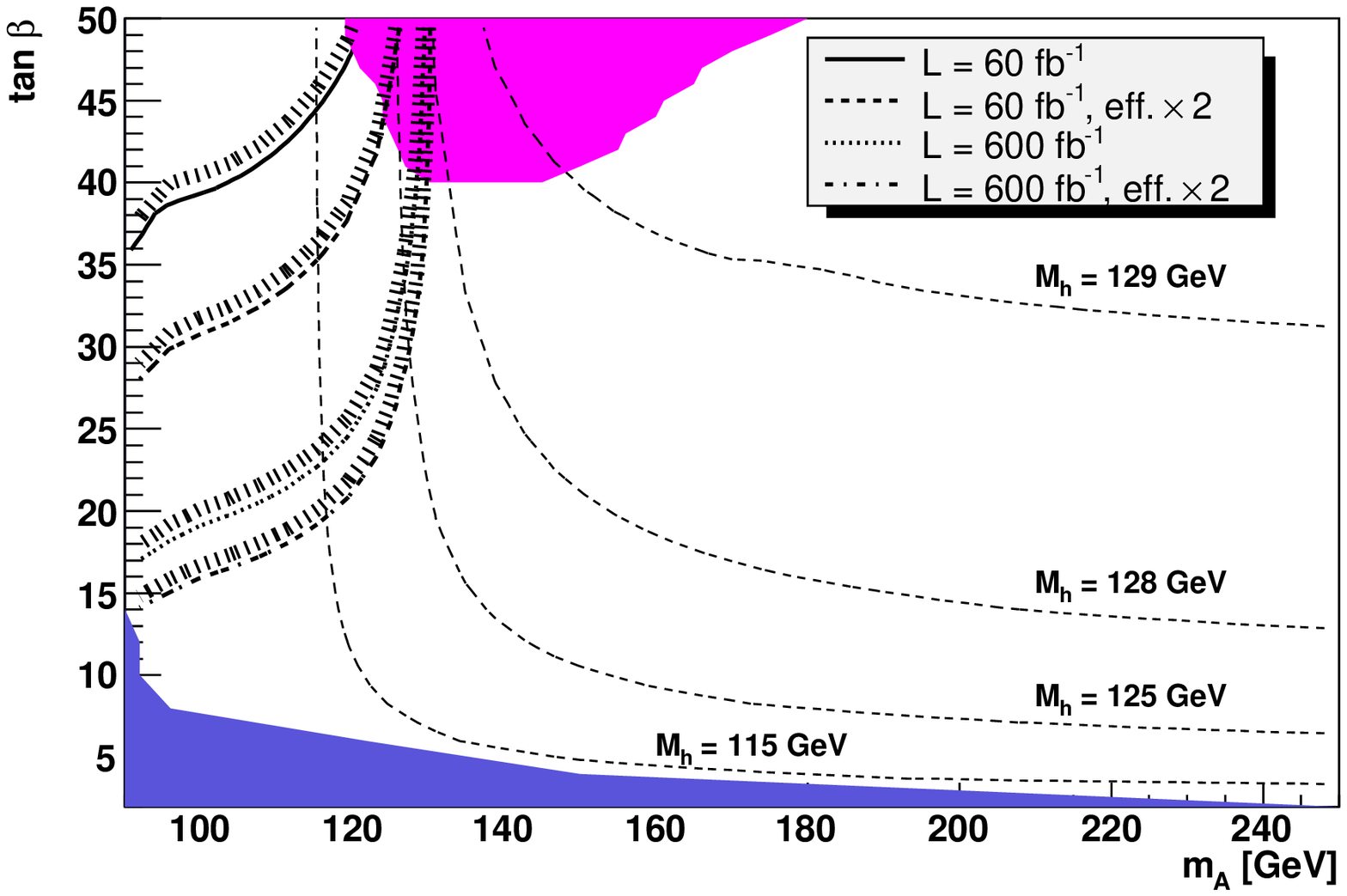}
\vspace{-2.0em}
\caption{Contours of
$3 \si$ statistical significance (left) 
and $5 \si$ discovery (right) contours for the $h \to b \bar b$ channel 
in the $\Mhmax$ benchmark scenario with $\mu = +200 \gev$. The results 
were calculated using 
Eqs.~(\ref{eq1}) and (\ref{eq:backbb}) for $A=0.92$ and $C=C_{\mathrm {NLO}}$
for effective luminosities of ``\sixoo'',
``\sixooeff'', ``\sixooo'' and ``\sixoooeff''.
The values of $\Mh$ are
shown by the contour lines. The medium dark shaded (blue)
regions correspond to the LEP exclusion bounds, while 
the Tevatron limits are  shown by the dark shaded (purple) regions.}
\label{fig:disc-h}
\end{figure}

The two plots in Fig.~\ref{fig:disc-h} exemplify our new
results for the case of $h$~production in the $\Mhmax$
scenario~\cite{benchmark}. They display the
contours of $3 \si$ statistical significance (left) and $5 \si$
discovery (right) in the $h \to b \bar b$ channel.
The left-hand plot shows that
while the allowed region at high $\tb$ and low $\MA$ can be probed also with 
lower integrated luminosity, in the ``\sixoooeff'' scenario the coverage
at the $3 \si$ level
extends over nearly the whole $\MA$--$\tb$ plane, with the exception of
a  window around $\MA \approx 130-140\gev$ (which widens up for small
values of $\tb$). The coverage
includes the case of a light SM-like Higgs, which corresponds
to the region of large $\MA$. It should be kept in mind that 
besides giving an access to 
the bottom Yukawa coupling, which is a crucial input for determining all
other Higgs couplings~\cite{HcoupLHCSM}, the
forward proton mode would
provide valuable information on the Higgs $\cp$ quantum numbers
and allow a precise Higgs mass measurement and maybe even 
a direct determination of its width.

\begin{figure}[htb!]
\vspace{-2em}
\includegraphics[width=.495\textwidth,height=4.5cm]
{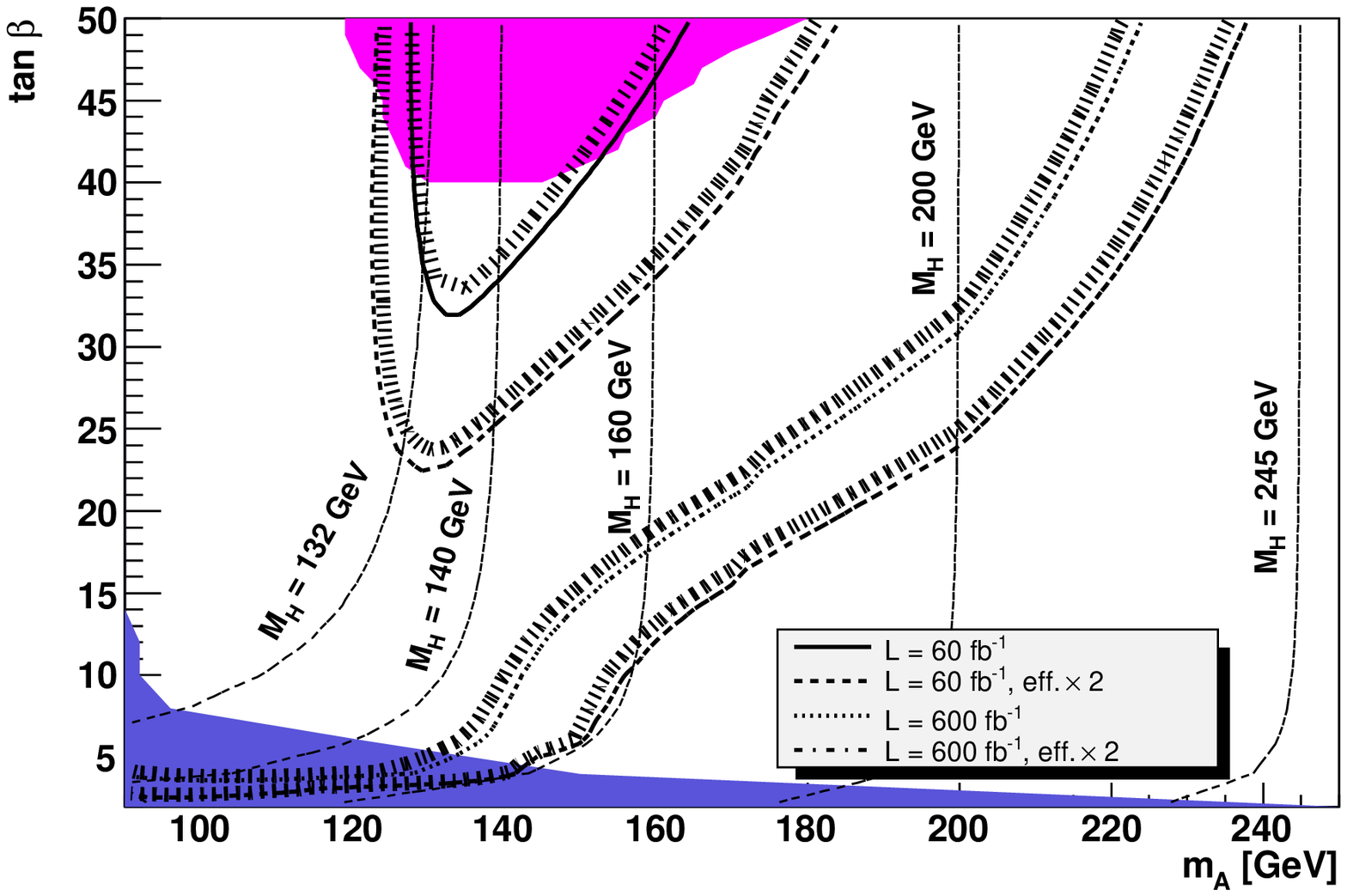}
\includegraphics[width=.495\textwidth,height=4.5cm]
{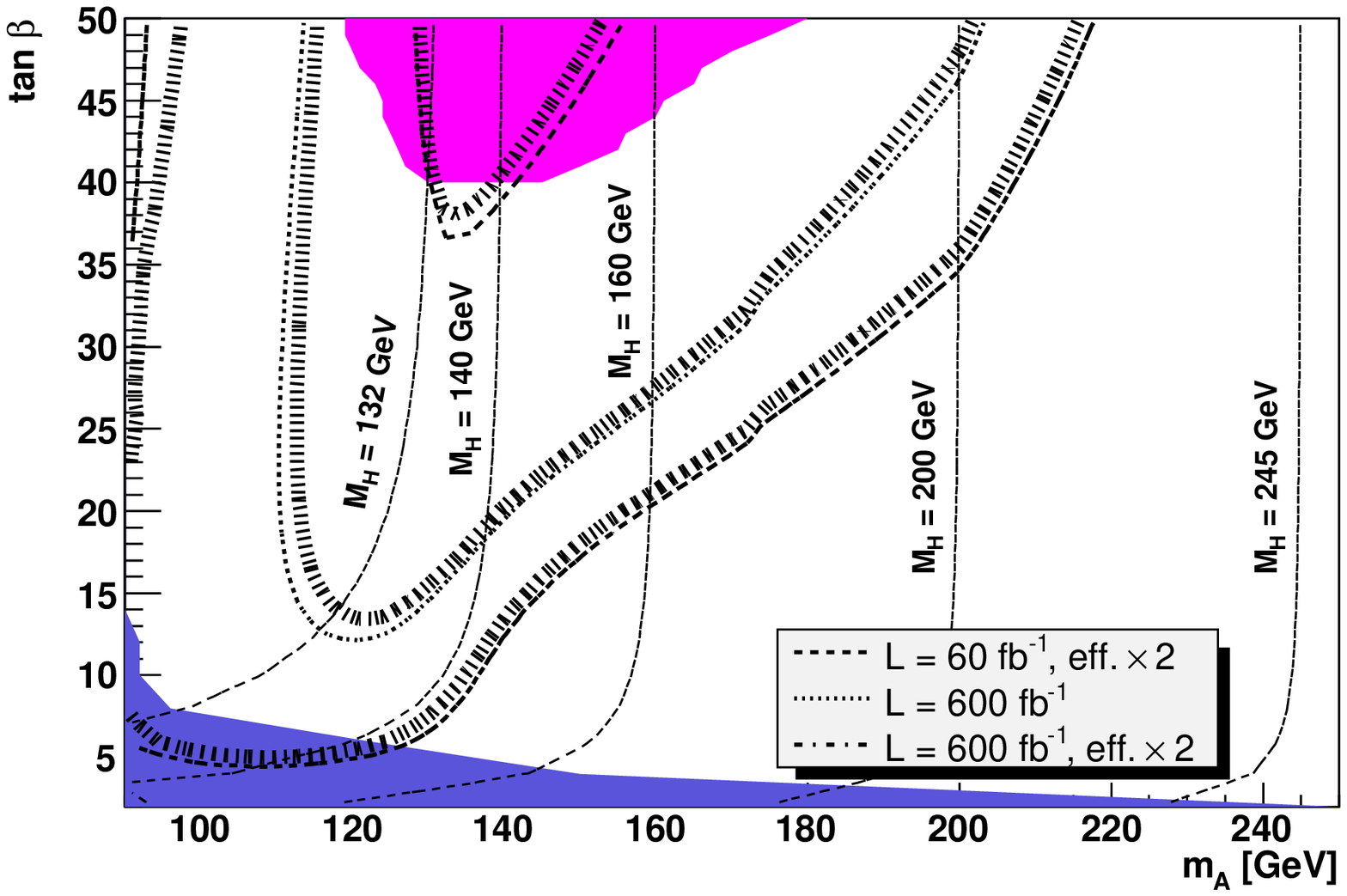}
\vspace{-2.0em}
\caption{
Contours of
$3 \si$ statistical significance (left) 
and $5 \si$ discovery (right) contours for the $H \to b \bar b$ channel, 
see Fig.~\ref{fig:disc-h}.}
\label{fig:disc-H}
\end{figure}

The properties of the  heavier boson $H$
differ very significantly from the ones of a SM Higgs with the same mass
in the region where $\MH \gsim 150 \gev$. While for a SM Higgs the
$\mbox{BR}(H \to b \bar b)$ is strongly suppressed, the
decay into bottom quarks is the dominant  mode for the MSSM Higgs boson~$H$.
The $3 \si$ significance and $5 \si$ discovery 
contours in the $\MA$--$\tb$ plane are displayed in the left and right
plot of Fig.~\ref{fig:disc-H}, respectively.
While the area covered in the ``\sixoo'' scenario is to a large extent
already ruled out by Tevatron Higgs searches~\cite{higgsbounds}, 
in the ``\sixoooeff'' scenario the reach for the
heavier Higgs at the $3\,\si$ level goes beyond $\MH \approx 235 \gev$ in
the large $\tb$ region. 
At the $5 \si$ level the reach is slightly reduced, but still 
extends beyond $\MH \approx 200 \gev$.
Thus, CED production of the $H$ with the subsequent decay to $b \bar b$ 
provides a unique opportunity for accessing its bottom Yukawa
coupling in a mass range where for a SM Higgs boson the $b \bar b$
decay rate  would be negligibly small.
In the ``\sixoooeff'' scenario the discovery of a heavy
$\cp$-even Higgs  with $M_H \approx 140 \gev$ will be possible for all
allowed values of $\tb$.



\smallskip
Concerning the determination of the spin and the CP properties of Higgs
bosons the standard methods rely to a large extent on the coupling of a
relatively heavy SM-like Higgs to two gauge bosons.
The first channel that should be mentioned here is $H \to ZZ \to 4l$.
This channel provides detailed information about spin and $\cp$-properties
if it is open~\cite{jakobs-rev}.

Within a SM-like set-up it was analyzed how the tensor structure of the 
coupling of the Higgs boson to weak gauge bosons can be determined at the 
LHC~\cite{HVV-LHC0,HVV-LHC2,HVV-LHC1}.
A study exploiting the difference in the
azimuthal angles of the two tagging jets in weak vector boson 
fusion has shown that for $\MHSM = 160 \gev$ 
the decay mode into a pair of $W$-bosons (which is maximal at 
$\MHSM = 160 \gev$) allows the discrimination between the two extreme
scenarios of a pure $\cp$-even (as in the SM) and a pure $\cp$-odd
tensor structure at a level of 4.5 to  5.3\,$\si$ using only about 10~\ifb 
(assuming the production rate is that of the SM, which is currently probed by
the Tevatron~\cite{TevHiggsSM}.)
A discriminating power of two standard deviations at $\MHSM = 120 \gev$
in the tau lepton decay mode requires an integrated luminosity of
30~\ifb~\cite{HVV-LHC1}. 

For $\MH \approx \MA \gsim 2 \MW$ the lightest MSSM
Higgs boson couples to gauge bosons with about SM strength, but its mass
is bounded from above by $\Mh \lsim 135 \gev$~\cite{feynhiggs},
i.e.\ the light Higgs stays below the threshold above which the decay to
$WW^({(*)}$ or $ZZ^{(*)}$ can be exploited.
On the other hand, the 
heavy MSSM Higgs bosons, $H$ and $A$, decouple from the gauge bosons. 
Consequently, the analysis for $\MHSM = 160 \gev$ cannot be taken over
to the MSSM. 
This shows the importance of channels to determine spin and
$\cp$-properties of the Higgs bosons without relying on (tree-level)
couplings of the Higgs bosons to gauge bosons.
CED Higgs production can yield crucial information in this
context~\cite{INC,KKMRext,HKRSTW}. 
The $\MHSM = 120 \gev$ analysis, on the other hand, 
can in principle be applied to the SUSY case. However, the coupling of
the SUSY Higgs bosons to tau leptons, in this case does not exhibit a
(sufficiently) strong enhancement as compared to the SM case, i.e.\ no
improvement over the $2 \si$~effect within the SM can be expected.
The same would be true in any other model of new physics
with a light SM-like Higgs and heavy Higgses that decouple from
the gauge bosons.


\section{Sensitivity to Higgs bosons in the SM4}

A very simple example of physics beyond the SM is a
model, ``SM4'', which extends the SM by a fourth generation of heavy
fermions, see, for instance, \cite{extra-gen-review}. 
In particular, the masses of the 4th generation quarks and leptons
are assumed to be (much) heavier than the mass of the top-quark.
In this case, the effective coupling of the Higgs boson
to two gluons is three times larger than in the SM.
No other coupling, relevant to LEP and Tevatron searches, changes
significantly.
Essentially, only the partial decay width
$\Gamma(H\to gg)$ changes by a factor of 9 and, with it, the 
total Higgs width and therefore all the decay branching ratios 
\cite{four-gen-and-Higgs}.
The new total decay width and the relevant decay branching ratios can be
evaluated as,
\newcommand{\TOT}{{\rm tot}}
\begin{align*}
\Gamma_\SM(H\to gg) & = \br_\SM(H\to gg)\:\Gamma_\TOT^\SM(H)\,,\\ 
\Gamma_\SMv(H\to gg) &= 9\:\Gamma_\SM(H\to gg)\,,\\
\Gamma_\TOT^\SMv(H) &= \Gamma_\TOT^\SM(H) - \Gamma_\SM(H\to gg)
	+ \Gamma_\SMv(H\to gg)\,.
\end{align*}
In \reffi{fig:SM4MH} we show the bounds on $\MHSMv$ from LEP and
Tevatron searches (taken from \cite{higgsbounds}, where also an
extensive list of experimental references can be found.)
Shown is the experimentally excluded cross section divided by the cross
section in the SM and the SM4, respectively.
The SM4 (SM) is given by the dashed (solid) line. In the 
red/light grey part the LEP exclusion provides the strongest bounds,
while for the blue/dark grey part the Tevatron yields stronger limits.
On can see that the exclusion bounds on $\MHSMv$ are much stronger than
on $\MHSM$, and only a window of $112 \gev \lsim \MHSMv \lsim 145 \gev$
is still allowed. At larger masses (not shown) 
$\MHSMv \gsim 220 \gev$ also remains
unexcluded. Consequently, we can focus our studies on the still allowed
regions. 

\begin{figure}[htb!]
\begin{minipage}[c]{0.55\textwidth}
\includegraphics[width=0.99\textwidth]{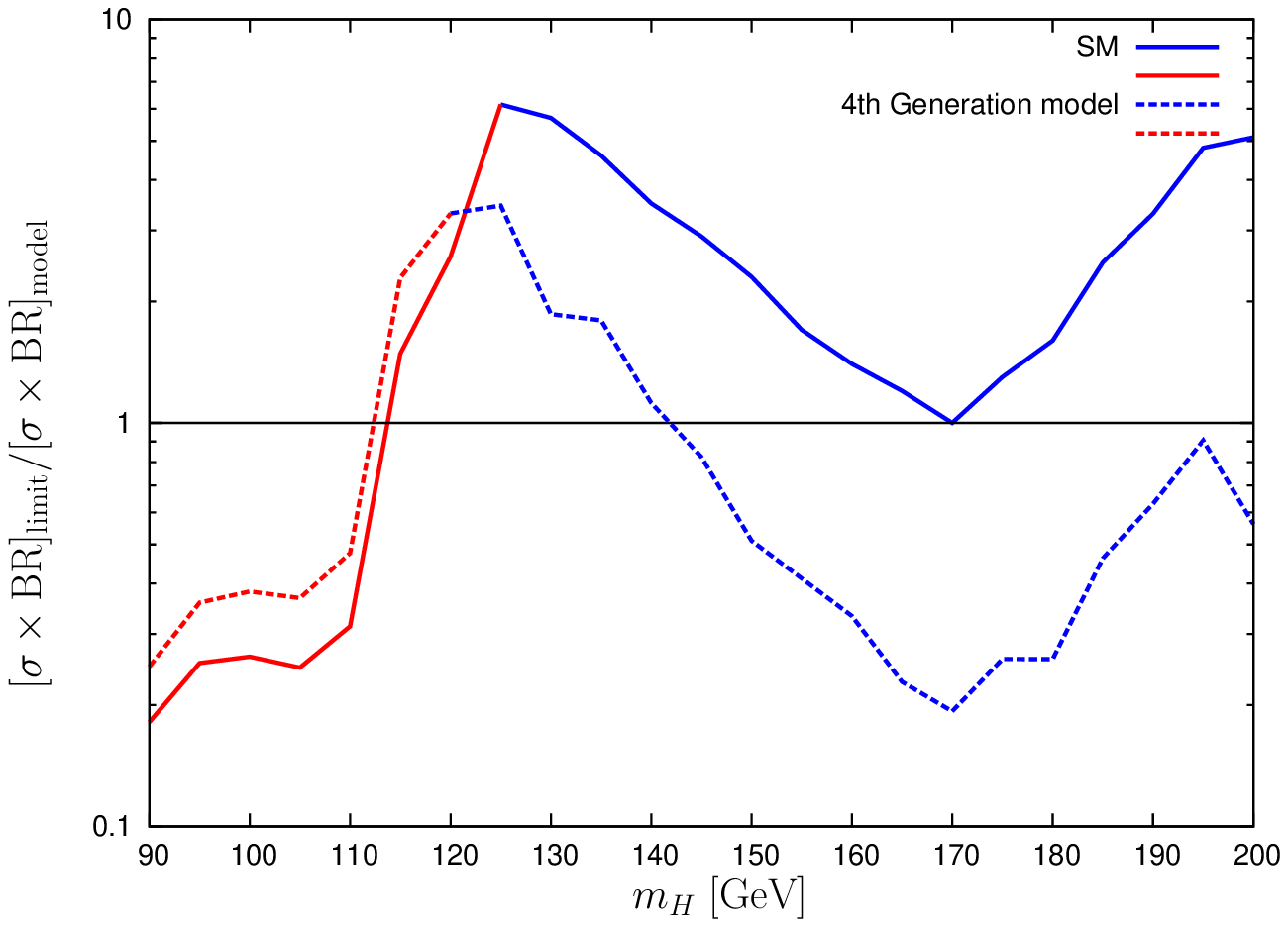}
\end{minipage}
\begin{minipage}[c]{0.03\textwidth}
$\phantom{0}$
\end{minipage}
\begin{minipage}[c]{0.40\textwidth}
    \caption{
	Cross section ratio (see text) for the most sensitive
	channel as a function of the Higgs mass $m_H$:
	4th Generation Model versus SM. 
	The colors indicate whether the most sensitive
	search channel is from 
	LEP (lighter grey) 
        or the Tevatron (darker grey), taken from~\cite{higgsbounds}.
        }
\label{fig:SM4MH} 
\end{minipage}
\end{figure}

\begin{figure}[htb!]
\vspace{-1.5em}
\includegraphics[width=.495\textwidth,height=5cm]
{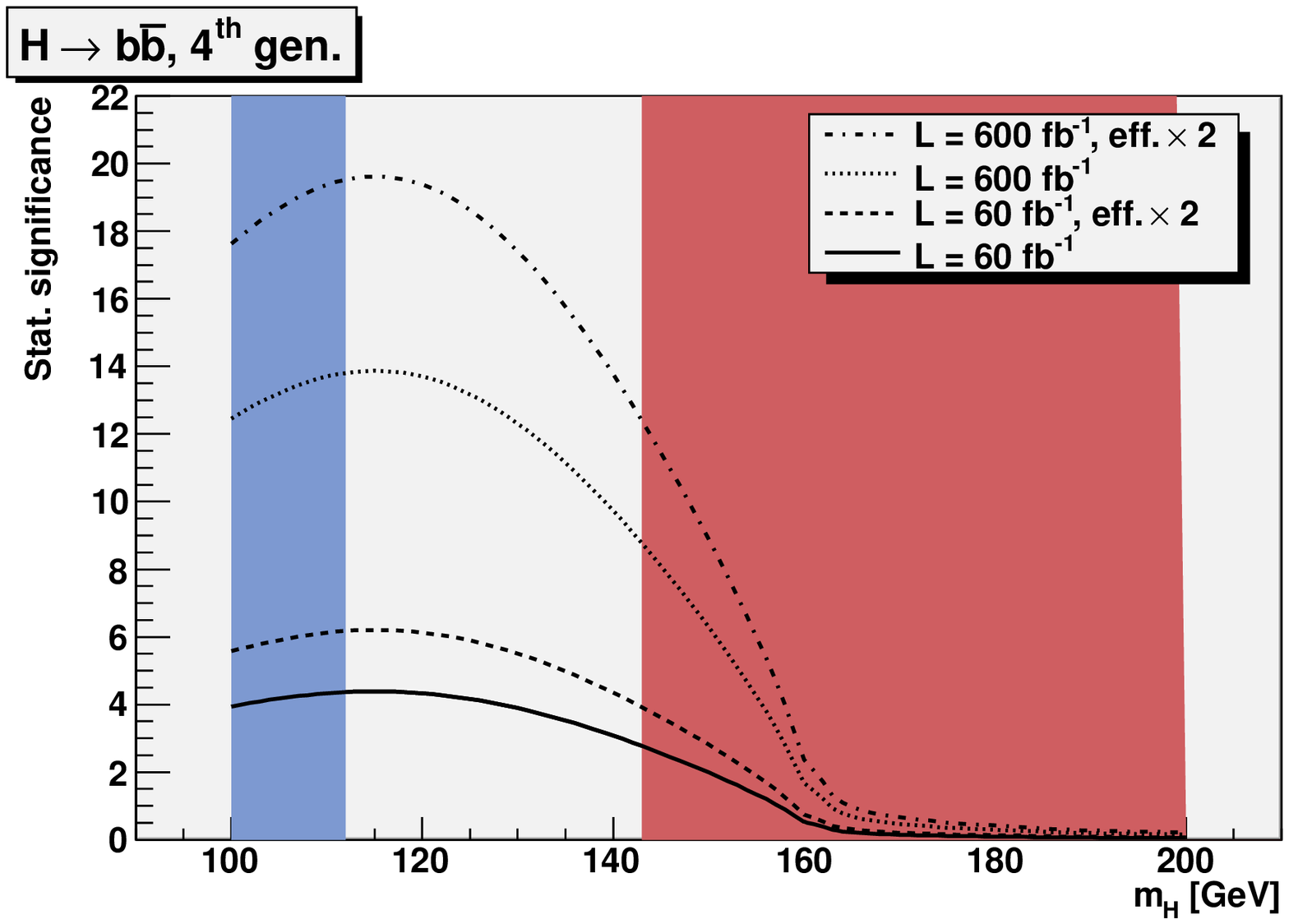}
\includegraphics[width=.495\textwidth,height=5cm]
{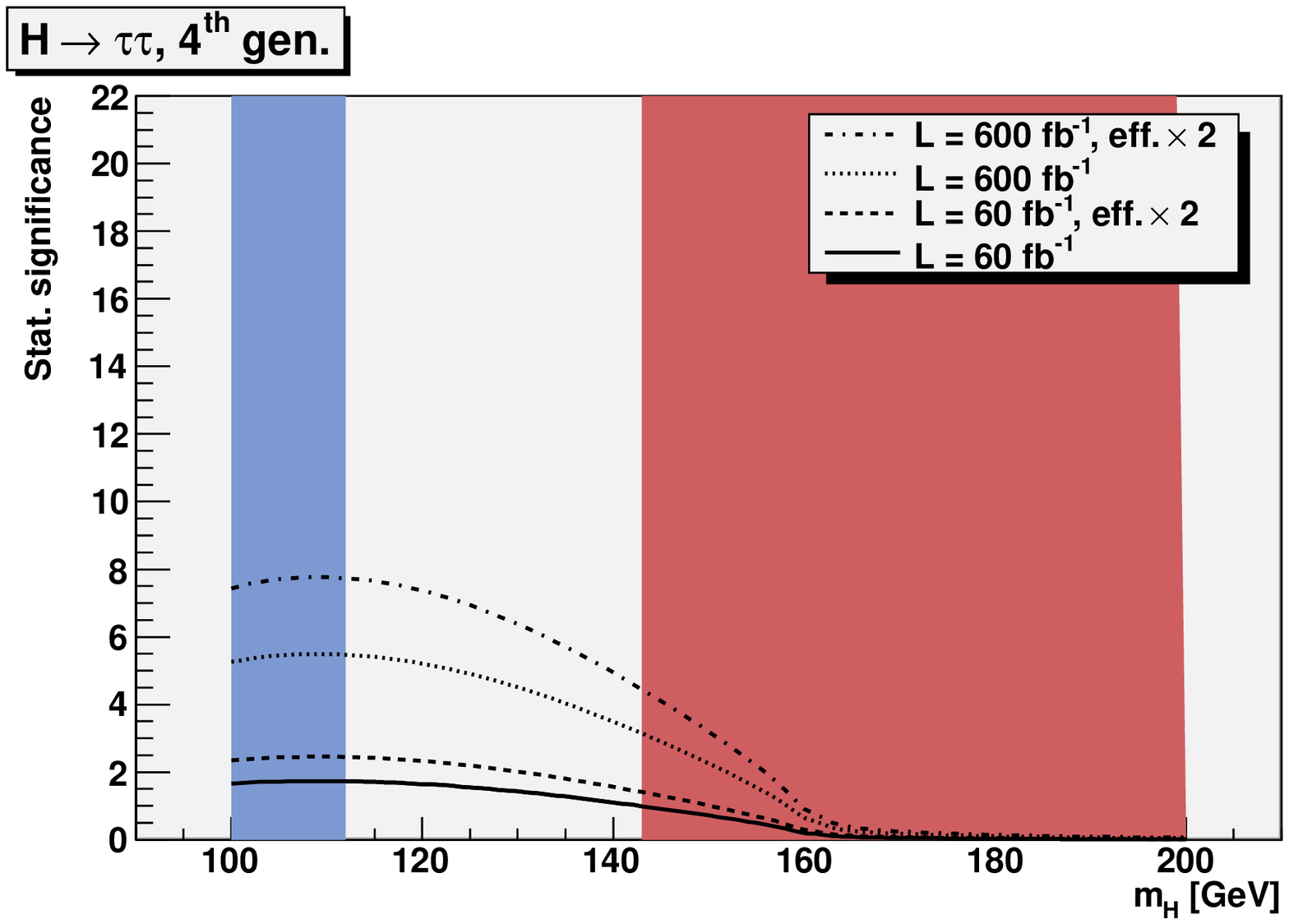}
\vspace{-1.5em}
\caption{Significances reachable in the SM4 in the $H \to b \bar b$ (left) and 
$H \to \tau^+\tau^-$ (right) channel 
for effective luminosities of ``\sixoo'',
``\sixooeff'', ``\sixooo'' and ``\sixoooeff''.
The regions excluded by LEP appear as blue/light grey for low values of 
$\MHSMv$ and excluded by the Tevatron as red/dark grey for larger values of
$\MHSMv$. 
}
\label{fig:disc-SM4}
\end{figure}

As for the MSSM we have evaluated the significances that can be obtained
in the channels $H \to b \bar b$ and $H \to \tau^+\tau^-$. The results
are shown in \reffi{fig:disc-SM4} as a function of $\MHSMv$ for the four
luminosity scenarios. 
The regions excluded by LEP appear as blue/light grey for low values of 
$\MHSMv$ and regions excluded by the Tevatron appears as red/dark grey
for larger values of $\MHSMv$. 
The $b \bar b$ channel (left plot) shows that even at rather low luminosity the
remaining window of $112 \gev \lsim \MHSMv \lsim 145 \gev$ can be
covered by CED Higgs production. Due to the smallness of 
$\br(\HSMv \to b \bar b)$ at $\MHSMv \gsim 160 \gev$, however, the CED
channel becomes irrelevant for the still allowed high values of
$\MHSMv$. The $\tau^+\tau^-$ channel (right plot) has not enough
sensitivity at low
luminosity, but might become feasible at high LHC luminosity.
At masses $\MHSMv \gsim 220 \gev$ it might be possible to exploit the decay 
$H \to WW, ZZ$, but no analysis has been performed up to now.


 

\begin{footnotesize}

\end{footnotesize}


\end{document}